\documentclass{article}
\PassOptionsToPackage{numbers, compress}{natbib}

\usepackage{graphicx}
\usepackage{subcaption}

\usepackage[final]{neurips_2022}

\usepackage[utf8]{inputenc} % allow utf-8 input
\usepackage[T1]{fontenc}    % use 8-bit T1 fonts
\usepackage{hyperref}       % hyperlinks
\usepackage{url}            % simple URL typesetting
\usepackage{booktabs}       % professional-quality tables
\usepackage{amsfonts}       % blackboard math symbols
\usepackage{nicefrac}       % compact symbols for 1/2, etc.
\usepackage{microtype}      % microtypography
\usepackage{xcolor}         % colors
\usepackage{amsmath}

\title{ViDa: Visualizing DNA hybridization trajectories with biophysics-informed deep graph embeddings}

\author{
  Chenwei Zhang \textsuperscript{1} \
  Jordan Lovrod \textsuperscript{1} \
  Boyan Beronov \textsuperscript{1} \
  Khanh Dao Duc \textsuperscript{2} \
  Anne Condon \textsuperscript{1} \\
  \textsuperscript{1} Department of Computer Science, UBC \;\;\;\;\;
  \textsuperscript{2} Department of Mathematics, UBC\\
  \texttt{\{cwzhang, jlovrod, beronov, condon\}@cs.ubc.ca} \;\;\;\;\; 
  \texttt{kdd@math.ubc.ca}\\
}

\begin{document}

\maketitle
\begin{abstract}
Visualization tools can help synthetic biologists and molecular programmers understand the complex reactive pathways of nucleic acid reactions, which can be designed for many potential applications and can be modelled using a continuous-time Markov chain (CTMC). Here we present ViDa, a new visualization approach for DNA reaction trajectories that uses a 2D embedding of the secondary structure state space underlying the CTMC model. To this end, we integrate a scattering transform of the secondary structure adjacency, a variational autoencoder, and a nonlinear dimensionality reduction method. We augment the training loss with domain-specific supervised terms that capture both thermodynamic and kinetic features. We assess ViDa on two well-studied DNA hybridization reactions. Our results demonstrate that the domain-specific features lead to significant quality improvements over the state-of-the-art in DNA state space visualization, successfully separating different folding pathways and thus providing useful insights into dominant reaction mechanisms.
\end{abstract}

\section{Introduction}

Nucleic acid nanotechnologies, including beacons \cite{Wang2009-nk}, riboswitches \cite{Roth2009-tb}, Boolean circuits \cite{qian-winfree-2011} and neural networks \cite{cherry-qian-2018}, are implemented using a series of reactions between multiple DNA or RNA strands. Molecular programmers would benefit from accurate estimates of the rates of such reactions, as they vary dramatically across sequences. Yet, the mechanisms that determine nucleic acid reaction kinetics are elusive, since they involve complex high dimensional trajectories over combinatorial spaces, i.e., sequences of secondary structure from the reactants to the products of a DNA reaction, along with the stochastic time spans taken to transition from one secondary structure to the next.
These secondary structures describe the set of base pairs formed via hydrogen bonding between Watson-Crick complementary bases, and each secondary structure has an associated free energy determined by latent thermodynamic parameters.
In this context, geometric deep learning methods \cite{graphconv,diffusioncnn,GSAE,bronstein2021geometric} provide a new potential strategy to represent energy landscapes of DNA and RNA secondary structure ``states'', as they have demonstrated success in analyzing graph-based data. 

In this paper, which significantly expands our original workshop paper \cite{Zhang-etal-2022}, we introduce a new workflow, called \textbf{ViDa}, for visualization of DNA reaction kinetics. Our approach uses deep graph embedding methods, augmented with biophysically informed features of the DNA reaction domain. Upon evaluating ViDa on two well-studied DNA reactions with different mechanisms, we demonstrate that ViDa's embeddings preserve both local structure, by clustering together states featuring similar motifs such as hairpins or stacks with common base pairs, as well as global structure, by keeping kinetically dissimilar regions of the state space far apart. Incorporating domain-specific features into the training of the neural embeddings appears to be critical to ViDa's success. Furthermore, the trajectories laid out smoothly on the 2D embedding reveal meaningful alternative folding pathways. Overall, these results suggest that ViDa can provide new mechanistic insights from sampled reaction trajectories.

\section{Related work} \label{relatedwork}

Elementary step simulators such as Multistrand \cite{multistrand} (see Appendix \ref{multistrand}) use CTMC models of reaction trajectories, and can stochastically generate trajectory samples. Multistrand's output uses ``dot-parenthesis'' (dp) notation to represent a secondary structure (see Appendix \ref{dp} and examples in Table \ref{table:samples}), and a sequence of such strings to represent structures along a trajectory. To situate trajectories in an energy landscape, Machinek et al. \cite{MachinekThreeway} used a coarse-grained map. However, the coarse-grained grid cells may include secondary structure states with very different free energies, making interpretation of different reaction trajectories difficult.

Castro et al. \cite{GSAE} developed a deep graph embedding framework, called the geometric scattering autoencoder (GSAE), to study energy landscapes of RNA secondary structures. GSAE has three major parts: an untrained geometric scattering transform \cite{scatteringtransform,zou2020graph,gama2018diffusion}, a trained variational autoencoder (VAE) \cite{VAE} and a trained auxiliary regression network, where the latter two networks together form a semi-supervised VAE. The geometric scattering transform ﬁrst extracts continuous high-dimensional features, called scattering coefﬁcients, from the discrete input graph, and these are then embedded into low-dimensional representations through the semi-supervised VAE. This embedding approximately retains important biophysical information, such as free energy, that can be used for further study. However, this approach is currently limited to single-stranded secondary structures, whereas many nucleic acid reactions of interest are typically multi-stranded, and it does not address the visualization of trajectories through such energy landscapes.

For further dimensionality reduction (DR) on the vector-valued VAE embedding, we apply PHATE (potential of heat diffusion for affinity-based transition embedding) \cite{PHATE}, a nonlinear and unsupervised DR method designed to capture both local and global structure among high-dimensional data points.

\section{Methods} \label{headings}

\subsection{ViDa workflow}
\begin{figure}[h]
  \centering
  \includegraphics[width=1\linewidth,trim={25cm 40cm 25cm 30cm} ,clip]
  {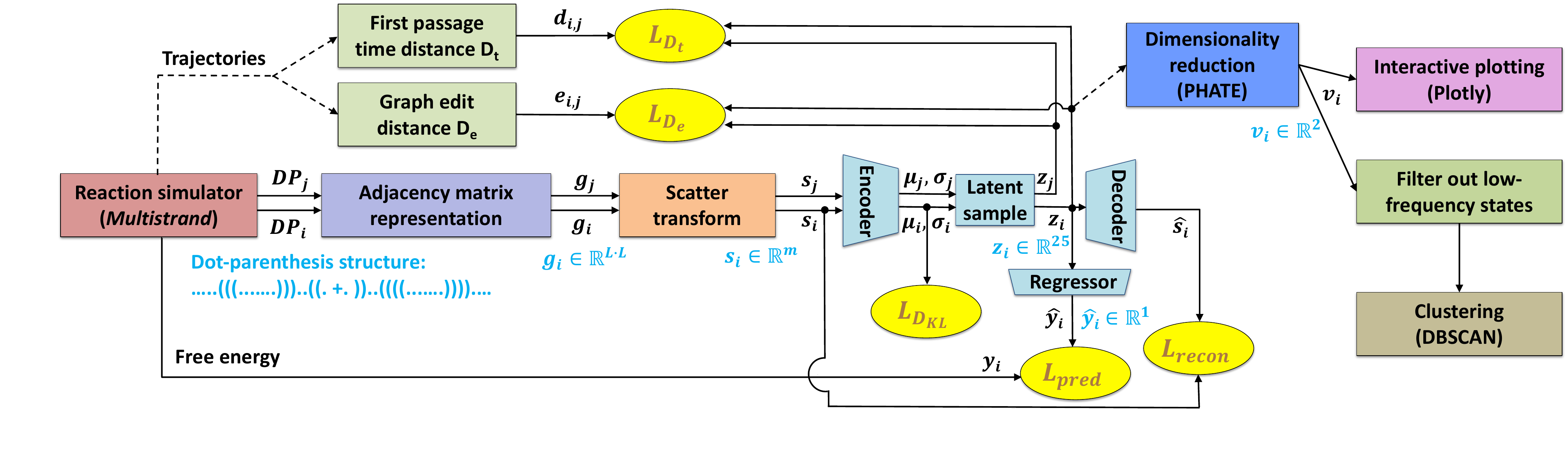}
  \caption{The ViDa framework consists of six major parts: the Multistrand reaction simulator \cite{multistrand}, a converter from dp notation to adjacency matrices, an untrained geometric scattering transform \cite{scatteringtransform,zou2020graph,gama2018diffusion}, a trained semi-supervised VAE \cite{VAE}, the nonlinear DR technique PHATE \cite{PHATE}, and post-processing components for interactive plotting and/or clustering. $DP_i$ is a sampled secondary structure, $g_i$ its graph adjacency representation and $s_i$ the corresponding vector of scattering coefficients, whereas 
  $\mu_i$ and $\sigma_i$ are the mean and standard deviation of the multivariate latent distribution,
  $z_i$ is a latent sample,
  $\hat{s}_i$ is the reconstructed scattering transform,
  $y_i$ and $\hat{y}_i$ are the simulated and regressed free energy values,
  $d_{i,j}$ and $e_{i,j}$ are the minimum passage time distance and graph edit distance between the secondary structures $i,j$,
  and $v_i$ is a 2D embedding of $z_i$.
  The training loss is composed of five terms: $L_{D_{KL}}$, $L_{recon}$ (unsupervised), $L_{pred}$, $L_{D_t}$, and $L_{D_e}$ (supervised).}
  \label{fig:ViDa}
\end{figure}

The ViDa framework pipeline is illustrated in Figure \ref{fig:ViDa}. Note that in this paper we only ultilize ViDa for double-stranded complexes, but it is also suitable for single-stranded structures such as hairpins.
An input set of secondary structures, represented using dp notation, their corresponding energies, as well as transition times between consecutively occupied states, were extracted from simulated Multistrand trajectories. 
Each state was converted to a graph adjacency matrix, with a node per nucleotide and two types of edges: strand backbones as determined by the primary structure, and complementary base pairs in the secondary structure. The resulting set of graphs ${G=\{g_1,g_2,...,g_n\}}$ was then passed through
a geometric scattering transform, which converts graph signals ${g_i \in \mathbb{R}^{L^2}}$ into scattering coefficient vectors ${s_i \in \mathbb{R}^{m}}$,
where $L$ is the sum of the lengths of the single-stranded sequences, $n$ is the total number of simulated states, and usually ${m > L^2}$.
Out of these coefficient vectors, 70\% were randomly assigned to the training set for the supervised VAE model, and the remaining 30\% were assigned to the testing set. 
The encoder network was comprised of two fully connected layers, followed by batch norm layers and RELU activations, and the decoder was chosen to be mirror symmetric.
In order to guide the training and to regularize the embedding space,
the latent samples produced by the encoder, ${z_i \in \mathbb{R}^d}$ with ${d \ll m}$, were additionally processed by a regressor network for predicting the free energy.
Overall, the VAE loss was augmented with regression terms for three domain-specific predictors:  the free energy, evaluated at each sampled $z_i$, as well as the ``minimum passage time'' distance $D_t$ (see Appendix \ref{mfpt}) and the graph edit distance $D_e$ for all pairs ${\left(z_i,z_j\right)}$.
Finally, the dataset ${Z_G=\{z_1,z_2,...,z_n\}}$ served as input to the DR algorithm PHATE, producing the 2D embedding $V_G=\{v_1,v_2,..,v_n\}, v_i \in \mathbb{R}^2$ for visualization and/or clustering.

\subsection{Domain-specific losses} \label{loss}
The total training loss for the ViDa model is made up of five terms:
\[
L_{tot} = \alpha L_{D_{KL}} + \beta L_{recon} + \gamma L_{pred} + \delta L_{D_t} + \epsilon L_{D_e}
\,.\]
The latent loss $L_{D_{KL}}$ and reconstruction loss $L_{recon}$ constitute the original VAE model, and we include
three domain-specific regression terms, namely the free energy loss $L_{pred}$ for the auxiliary regression network, the minimum passage time distance loss $L_{D_t}$, and the graph edit distance loss $L_{D_e}$. The free energy loss is calculated using the ground truth values $y_i$ from the Multistrand simulator,
\begin{equation}
    L_{pred}= \sum_{i}\left(\hat{y}_i - y_i\right)^2 \,,
\end{equation}
and we define the graph edit distance loss as
\begin{equation}
    \begin{split}
    L_{D_e}= \sum_{i,j}{\left(||z_i - z_j|| - e_{i,j}\right)^2} \,,
    \end{split}
\end{equation}
where $e_{i,j}$ is the graph edit distance from state $i$ to $j$.
Since we have converted secondary structures to adjacency matrices at an early stage before training, it is convenient to compute the graph edit distance between two states by simply subtracting their corresponding adjacency matrices.
Analogously,
we define the minimum passage time distance loss as
\begin{equation}
    \begin{split}
    L_{D_t}= \sum_{i,j}w_{i,j}\cdot{\left(||z_i - z_j|| - d_{i,j}\right)^2} \,,
    \end{split}
\end{equation}
where $d_{i,j}$ is an estimate of the minimum passage time from $i$ to $j$ or from $j$ to $i$, computed from the simulated trajectories as explained in Appendix \ref{mfpt}. Here,
\begin{equation}
    \begin{split}
    w_{i,j}=p_ip_j\cdot\mathbb{I}[d_{i,j}\leq\bar{d}]
    \end{split}
\end{equation}
is an importance weight, based on the empirical probability $p_i$ of state $i$ as well as the indicator function $\mathbb{I}[\dots]$, which limits the loss to state pairs with minimum passage time bounded by a given threshold $\bar{d}$.

\subsection{Implementation}
The VAE is intended to be trained separately for each DNA reaction. For our experiments,
the bottleneck dimension of the VAE was set to $d=25$ and training was performed using PyTorch’s Adam optimizer.
The maximum epoch size was set to $150$ to avoid overﬁtting with a batch size of $64$,
where the initial learning rate was set to $0.0001$ and then dynamically adjusted by the ReduceLROnPlateau scheduler with default parameters, except for ${\texttt{patience}=5}$. 
The hyperparameters for the VAE loss were set to $\alpha=1$, $\beta=0.0001$, $\gamma=0.3$, $\delta=0.0001$, and $\epsilon=0.0001$ in all cases. 
For PHATE, the number of landmarks was set to $2000$, the decay rate to $40$ and the number of nearest neighbours to $5$. 
The interactive plotting tool used the Plotly library. The clustering method used DBSCAN (density-based spatial clustering of applications with noise) \cite{DBSCAN} with parameters $\texttt{eps}=0.005$ and $\texttt{min\_samples}=4$ (see Appendix \ref{DBSCAN}).

\section{Results} \label{results}

In this section, we present and assess ViDa's visualizations of two DNA \textit{hybridization} reactions, wherein two unbound complementary strands bind and fold into a double-stranded helix. The first reaction, which we denote by Gao-P4T4, is from Gao et al.'s experimental study \cite{gaohelix}, and the second, which we denote by Hata-39, is from Hata et al.'s experimental study \cite{hata}. The sequences for the two reactions are shown in Table \ref{table:samples}, along with some key possible secondary structure motifs for each. For visualizations of Gao-P4T4 obtained using other DR approaches, including PCA, PHATE, GSAE+PCA, GSAE+PHATE, and MDS (multidimensional scaling), see Appendix \ref{otherviz}. 

\begin{table}[h]
  \caption{Sequences of reactions Gao-P4T4 \cite{gaohelix} and Hata-39 \cite{hata}, and examples of key sequence-dependent secondary structure motifs that affect their reactive pathways and reaction rate. \\ }
%   \vspace{1cm}
  \label{table:samples}
  \small
  \centering
  \begin{tabular}{ll}
    \toprule
   \multicolumn{2}{l}{\textbf{Dp notation for the reactants and products of any standard hybridization reaction} (e.g. 25 bases per strand)} \\
    unbound structure: & 3$'$-\texttt{.........................}-5$'$ + 3$'$-\texttt{.........................}-5$'$ \\
    hybridized structure: & 3$'$-\texttt{(((((((((((((((((((((((((}-5$'$ + 3$'$-\texttt{)))))))))))))))))))))))))}-5$'$ \\
    \midrule
   \multicolumn{2}{l}{\textbf{Gao-P4T4} (25 bases per strand)} \\
   sequences: & 3$'$-\texttt{ACACGATCATGTCTGCGTGACTAGA}-5$'$ + 3$'$-\texttt{TCTAGTCACGCAGACATGATCGTGT}-5$'$  \\
    possible hairpins (size 3): & 3$'$-\texttt{..........(((.....)))....}-5$'$ + 3$'$-\texttt{....(((.....)))..........}-5$'$ \\
    possible hairpins (size 4): & 3$'$-\texttt{.((((..........))))......}-5$'$ + 3$'$-\texttt{.....((((...........)))).}-5$'$ \\
    \midrule
   \multicolumn{2}{l}{\textbf{Hata-39} (23 bases per strand)} \\
    sequences: & 3$'$-\texttt{CCATCAGGAATGACACACACAAA}-5$'$ + 3$'$-\texttt{TTTGTGTGTGTCATTCCTGATGG}-5$'$  \\
    possible hairpin (size 3): & 3$'$-\texttt{.(((.....)))...........}-5$'$ + 3$'$-\texttt{.......................}-5$'$ \\
    possible mis-stack (size 7): & 3$'$-\texttt{..............(((((((..}-5$'$ + 3$'$-\texttt{....)))))))............}-5$'$ \\
    \bottomrule
  \end{tabular}
\end{table}

\subsection{Case study 1: Gao-P4T4}

The strands in Gao-P4T4, which involve 25 bases each, were designed such that 4-stem hairpins could form \cite{gaohelix}. The experimental hybridization measurements from this study are currently best understood with the follow-up analyses by Schreck et al. \cite{schreck}. They argue that the 4-stem hairpins slow hybridization primarily by destabilizing partially formed duplexes, rather than by occluding potential binding sites or impeding the ``zippering'' of strands. 

For our visualization study, we generated 100 trajectory samples using Multistrand's \textit{trajectory mode} (see Appendix \ref{multistrand}). The initial state for our simulations was the unbound structure with no base pairs, and the final state was the fully hybridized structure in which all bases are paired to their intended complement (see Table \ref{table:samples}). All 46606 unique states found during simulation are included in the embedding.
In the plots of this subsection, states are coloured according to their free energy, and some of the arguments about the quality of our embedding rely on the energy trends observed in the plots.

\paragraph{ViDa preserves global and local structure in energy landscapes.} \label{finding1}

\begin{figure}[h]
    \centering
    \begin{subfigure}[b]{0.5\textwidth}
        \includegraphics[width=1\linewidth, trim={0cm 8cm 17.5cm 1cm} ,clip] 
        {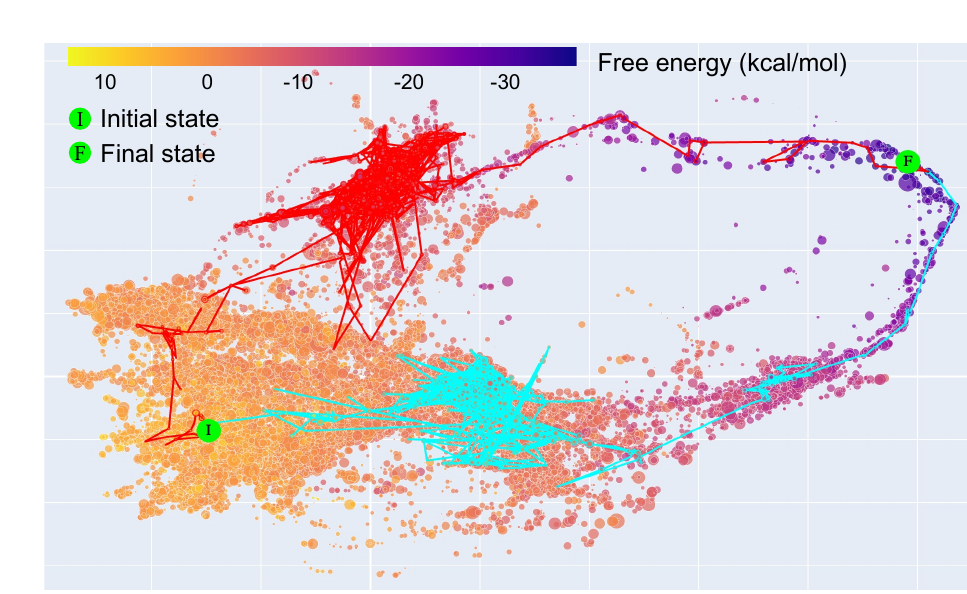}
        \caption{Two trajectories over ViDa embedding  \label{fig:vida-pt4}}
    \end{subfigure}
    \begin{subfigure}[b]{0.49\textwidth}
        \includegraphics[width=1\linewidth, trim={17.7cm 8cm 0cm 1cm} ,clip]
        {Figures/PT4.pdf}
        \caption{DBSCAN clusters in ViDa embedding \label{fig:dbscan-pt4}}
    \end{subfigure}\hfil
    \caption{
    ViDa embedding results for Gao-P4T4. Each point represents a secondary structure state. The green circle marked $I$ ($F$) denotes the initial (final) state.
    \textbf{(a)} 2D embedding of secondary structure states. The colour of each state refers to its free energy. The diameter of each state is proportional to its average sampled holding time. The red and cyan traces represent two different trajectory samples. 
    \textbf{(b)} Results of DBSCAN on the 2D embedded states. DBSCAN identifies three clusters (cyan, red, and dark grey). The stars indicate the minimum free energy state within each cluster. States that are not clustered are shown in light grey.}
    \label{fig:PT4}
\end{figure}

The secondary structure embedding for Gao-P4T4 is shown in Figure \ref{fig:vida-pt4}. The free energy, which is superimposed on the embedding plot, follows a high-to-low trend from the unbound (initial) state to the hybridized (final) state, suggesting that ViDa preserves global structure. Furthermore, by manually hovering over the points in the interactive plot, we find that neighbouring structures often only differ by a few base pairs, suggesting that ViDa also preserves local structure.

\paragraph{ViDa provides a more nuanced understanding of reaction mechanisms.} \label{finding3}

The embedding (Figure \ref{fig:vida-pt4}) separates states into two main branches. The lower branch corresponds to the reactive pathway in which the helix begins forming at the 5$'$ end of the first strand. In other words, this branch contains most structures of the form {3$'$-{[$\ldots$](((}-5$'$+3$'$-{)))[$\ldots$]}-5$'$}, where the [$\ldots$]s together comprise a legal dp sub-structure. These structures often coincide with the formation of the stable 4-stem hairpins with large loops (see Table \ref{table:samples} and Figure \ref{fig:4stem}), and is therefore a slow reactive pathway. On the other hand, the top branch corresponds to the reactive pathway in which inter-strand base pairs form at the 3$'$ end of the first strand (structures of the form {3$'$-{((([$\ldots$]}-5$'$+3$'$-{[$\ldots$])))}-5$'$}), and small 3-stem hairpin often form in both strands (see Table \ref{table:samples}). 
Laying out trajectories on the embedding, we find two dense regions (see Figure \ref{fig:vida-pt4}). We hypothesized the presence of kinetic traps within these regions. To delve deeper into these regions of interest, we first excluded less significant states with exceedingly low empirical state probabilities and then employed DBSCAN to cluster the post-filtered states (see Figure \ref{fig:dbscan-pt4}). We obtained three clusters, that also each locate around states with minimum free energy (MFE). Upon investigating these traps, we first found that for the kinetic trap in the cyan cluster, its corresponding secondary structure is {3$'$-{.((((..........)))).((((.}-5$'$+3$'$-{.)))).((((..........)))).}-5$'$}. The 4-stem hairpins that are extremely stable and hard to break have the same structures as the design of Gao-P4T4 (see Table \ref{table:samples}), providing a barrier to hybridization that is consistent with the computational analysis by Schreck et al. \cite{schreck}. The second kinetic trap we investigated (in the grey area) has a secondary structure of {3$'$-{............((((((((((((.}-5$'$+3$'$-{.)))))))))))).(((....))).}-5$'$}. The presence of a solitary 3-stem hairpin in one strand could be viewed as a minor trap due to its relatively poor stability. Finally, the kinetic trap in the third cluster (red) has a secondary structure of {3$'$-{.(((((((..(((.....)))...}-5$'$+3$'$-{...(((.....)))..))))))).}-5$'$}, with two 3-stem hairpins at both strands (see Figure \ref{fig:3stem}). These two 3-stem hairpins are more stable than the solo one, and impede the hybridization process, thus slowing down the overall reaction. 
In summary, our visualization highlights the kinetic trap created by the designed 4-stem hairpins in Gao-P4T4 reaction, which is stable enough to significantly slow down hybridization. Additionally, we identified a second major kinetic trap in Gao-P4T4, with 3-stem hairpins on both strands, which exacerbates the slowness of the reaction process.

\paragraph{Domain-specific features improve trajectory smoothness.} \label{finding2}

Laying out the trajectories on the embedding, all trajectories proceed nicely along the branches (Figure \ref{fig:vida-pt4}). Additionally, for all trajectory plots, we did not observe large jumps occurring along the traces, confirming that nearby secondary structures on simulated trajectories tend to be placed nearby in the embeddings. 
In order to quantify this smoothness property, we use a custom metric for distortion/stretch. We define the \emph{average distortion} of an embedding as the frequency-weighted mean Euclidean distance between the images of secondary structure pairs that occur consecutively in the trajectory dataset, normalized by the embedding diameter of all states. In Table \ref{table:smoothness} we compare the average distortion achieved by ViDa and by general-purpose DR methods (GSAE, MDS, PHATE, and PCA), and find that our model achieves a significantly lower distortion than all other considered methods.
On the one hand, PHATE and PCA do not take into account any domain knowledge beyond the training data itself, and thereby their visualizations and smoothness are both relatively poor. The comparisons suggest the importance of incorporating domain-specific knowledge when training neural networks to make a biophysically-plausible visualization tool, such as our custom loss terms
(Section \ref{loss}) that penalize the distortion of local structure. 
On the other hand, we also compared an MDS embedding 
which only leverages the biophysics-based distance measure of minimum passage time (see Figure \ref{fig:mds}). However, trajectories are densely concentrated around the initial state in the embedding, making it infeasible to distinguish different folding pathways.
These results emphasize the significance of integrating deep graph embeddings and distance loss metrics for achieving superior results.
In combination with the visualizations, they demonstrate that ViDa can embed the reaction trajectories while preserving some continuity in time.

\begin{table}
    \captionsetup{skip=5pt}
    \caption{Comparison of average distortion for different embedding methods for Gao-P4T4.}
    \label{table:smoothness}
    \centering
    \begin{tabular}{l|cccccc}
        \toprule
        \cmidrule{1-7}
        Metric  & ViDa (ours)  & PCA & PHATE  & MDS & GSAE+PCA  & GSAE+PHATE  \\
        \midrule
        \emph{Avg. distortion} & $\mathbf{0.019}$  &  $0.159$   &  $0.105$ &  $0.081$  &  $0.035$  & $0.030$  \\
        \bottomrule
    \end{tabular}
\end{table}

\paragraph {Comparison with state-of-the-art coarse-grained visualizations for hybridization.} \label{finding4}

\begin{figure}[h]
  \centering
  \includegraphics[width=0.7\linewidth,trim={2cm 4cm 6.5cm 2.5cm} ,clip]
  {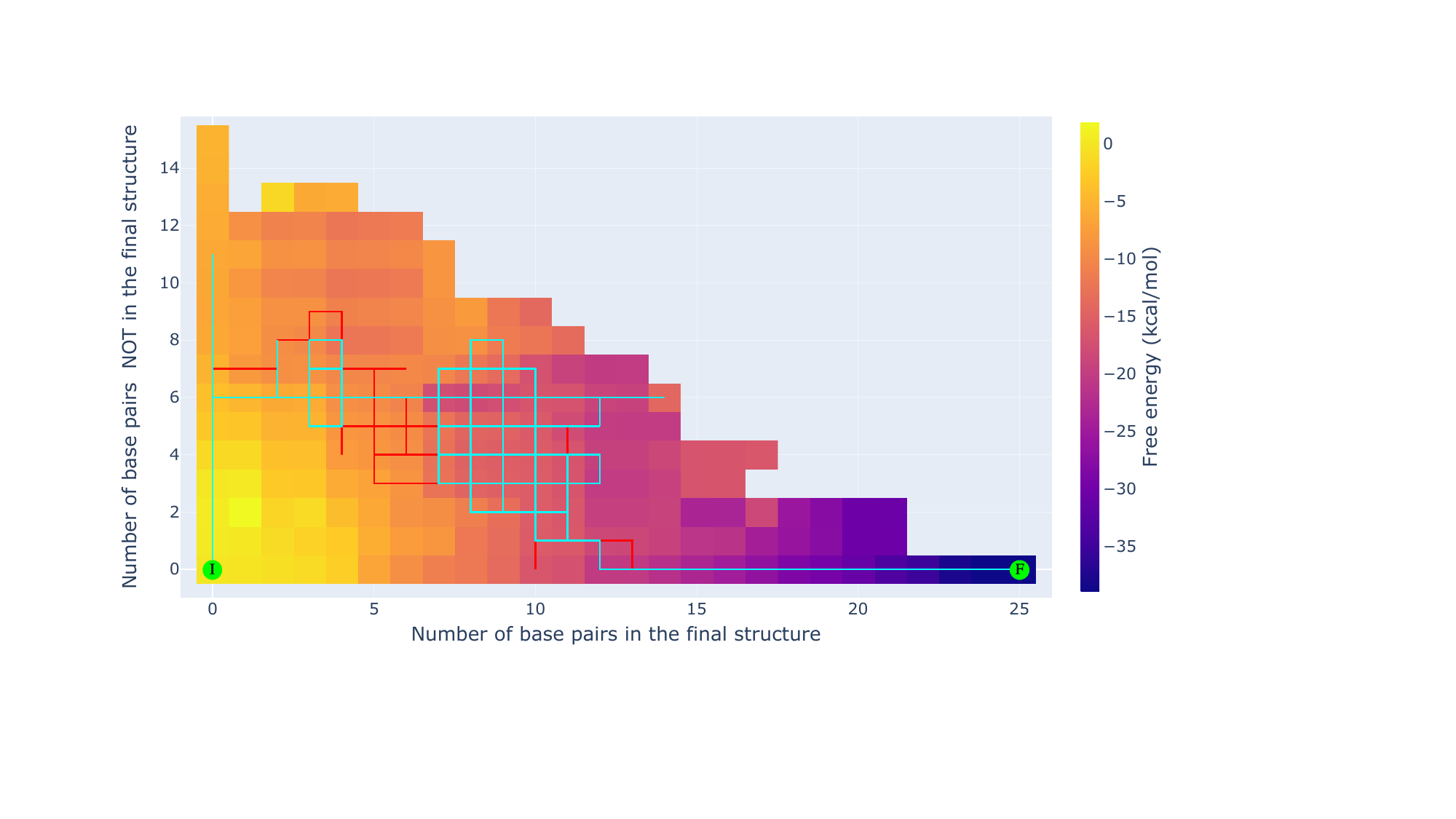}
  \caption{Coarse-grained visualization of Gao-P4T4. Each grid cell is an ensemble of secondary structures. A structure is in cell $(x,y)$ exactly $x$ of its base pairs contribute to the desired helix, and exactly $y$ of its base pairs are {\em not} part of the desired helix. The initial state (denoted by the green circle marked $I$) corresponds to the grid cell (0,0) and the final state (denoted by the green circle marked $F$) corresponds to the grid cell (25,0), since the strands have 25 bases each. The red and cyan traces are the same two trajectory samples shown in Figure \ref{fig:vida-pt4}.
  }
  \label{fig:pt4cg}
\end{figure}

In Figure \ref{fig:pt4cg}, we show a coarse-grained representation of Gao-P4T4, similar to visualizations in \cite{MachinekThreeway,PE}. Each secondary structure is mapped to a single macrostate based on (1) the number of base pairs that correspond exactly to base pairs in the desired helix and (2) the number of base pairs that do \textit{not} contribute to the desired helix, for instance base pairs involved in hairpins or mis-stacks. Each macrostate is therefore an ensemble of secondary structures. These sorts of coarse-grained visualizations are easily adjustable, do not require training, and have the capacity to represent all possible secondary structure states and trajectories. However, with this scheme, structurally dissimilar secondary structures may be mapped to the same macrostate, making it difficult to interpret each macrostate and trajectories through them, and to distinguish between different reaction mechanisms. 
In contrast, ViDa's fine-grained embedding overcomes this limitation. ViDa's plots show distinct reaction trajectories, enabling users to interpret reaction mechanisms more straightforwardly and accurately.

\subsection{Case study 2: Hata-39} \label{finding5}

The strands in Hata-39, which involve 23 bases each, were designed with the intention of making mis-nucleation and hairpin formation unlikely \cite{hata}. Hata-39 is currently best understood with the follow-up analyses by Lovrod et al. \cite{lovrod}. They show that the Hata-39 sequence gives rise to important secondary structures that are not common among hybridization reactions, and not generally considered in hybridization models. More specifically, it is possible for these strands to form stable stacks (3+ consecutive desired inter-strand base pairs), stable mis-stacks (3+ consecutive \textit{undesired} inter-strand base pairs), and hairpins (of size 3+) \textit{simultaneously}, leading to a diverse set of reactive pathways. The analysis involves a definition of eight \textit{structural types} of secondary structures, which we use in this subsection to colour the states in each plot and argue about the quality of our embedding. 

For our visualization study, we use 50 reactive pathway samples and 3095 non-reactive pathway samples that were generated using Multistrand's \textit{first step mode} (see Appendix \ref{multistrand}). In each first step mode simulation, an initial state is Boltzmann sampled from the set of all structures with exactly one inter-strand base pair, and the simulation is stopped when the two strands unbind, or when all bases are paired to their intended complement. All 56702 unique states found during simulation are included in our embedding.

\begin{figure}[h]
    \centering
    \begin{subfigure}[b]{0.5\textwidth}
        \includegraphics[width=1\linewidth, trim={0cm 7.5cm 17.4cm 1cm} ,clip] 
        {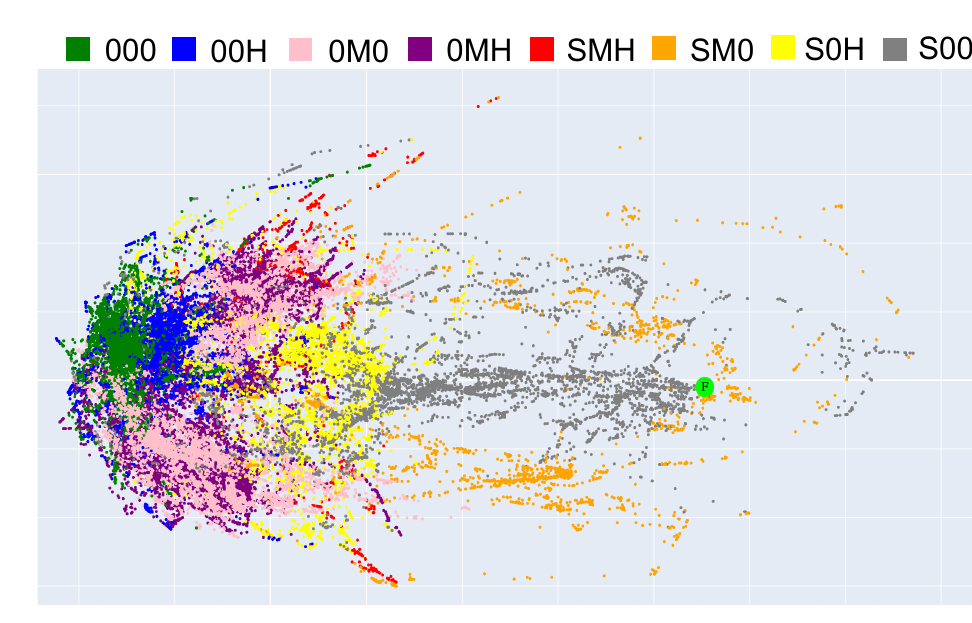}
        \caption{{ViDa embedding}
        \label{fig:vida-hata39}}              
    \end{subfigure}
    \begin{subfigure}[b]{0.488\textwidth}
        \includegraphics[width=1\linewidth, trim={17.8cm 7.5cm 0cm 1cm} ,clip]
        {Figures/Hata39.pdf}
        \caption{{Three trajectories over ViDa embedding}
        \label{fig:hatatrj}}
    \end{subfigure}\hfil
    \caption{ViDa embedding results for Hata-39. Each point represents a secondary structure state. The green circle marked $F$ denotes the final (hybridized) state. 
    \textbf{(a)} 2D embedding of secondary structure states. The colour of each state refers to its structural type \cite{lovrod}.  For instance, \textit{SM0} denotes the type of secondary structure states with at least one stack, at least one mis-stack, and no hairpins. There are eight structural types in total. 
    \textbf{(b)} Three reactive trajectories laid over the embedding. States that do not lie on one of the three trajectories are shown in light grey.}
    \label{fig:Hata39}
\end{figure}

\paragraph{ViDa embedding is compatible with structural types.} The secondary structure embedding for Hata-39 is shown in Figure \ref{fig:vida-hata39}. To assess the quality of the embedding and establish compatibility with previous work, we colour each state according to its structural type, which is determined by whether there is at least one correctly hybridized stack (\textbf{S}) or not (\textbf{0}), at least one mis-stack (\textbf{M}) or not (\textbf{0}), and at least one hairpin (\textbf{H}) or not (\textbf{0}) \cite{lovrod}. Schematic representations for stacks, mis-stacks, and hairpins are given in Figure \ref{fig:hairpins}. 
Although training ViDa does not receive these structural labels as input, states with the same, or similar, types tend to be close together in the embedding, implying that ViDa captures local structure. For instance, states of type \textit{0M0} (pink) are nearby states of type \textit{0MH} (purple), and indeed these structures are closely related because they contain a similar mis-stack, which dominates the hairpins in this reaction (see Table \ref{table:samples}). Moreover, states with very different structures are far apart in the embedding. For instance, the states of type \textit{00H} (blue) are generally far from states of type \textit{SM0} (orange), which is reasonable since they don't share any significant structural motifs stacks, mis-stacks, or hairpins. This provides evidence that the embedding preserves global structure.

\paragraph{ViDa can distinguish between different reaction mechanisms.} \label{finding6}

The structural labels can also highlight the reaction mechanisms captured by the embedding. Figure \ref{fig:hatatrj} shows three trajectory samples, which are representative of three distinct reaction mechanisms, laid over the embedded secondary structures. The black trajectory is an example of a direct hybridization reaction mechanism, which is extremely fast, but only accounts for $\sim$10\% of the sampled reactive trajectories. The orange and purple trajectories illustrate slower, more complex reactive pathways that, in the case of this reaction, are much more common. Similar to the dominant mechanism described for Gao-P4T4, the orange trajectory includes the formation of a 3-stem hairpin, such as 
{3$'$-{.(((.....))).((((((((..}-5$'$+3$'$-{..))))))))..((.....))..}-5$'$}, 
making this pathway slower than the direct pathway. The purple trajectory, although it also involves the formation of a 3-stem hairpin, is qualitatively distinct from the other two trajectories in its formation of a stable mis-stack, e.g. 
{3$'$-{.(((.....)))..(((((((..}-5$'$+3$'$-{....)))))))...........}.-5$'$}. 
These three reaction mechanisms, originally found and illustrated by Lovrod et al. \cite{lovrod}, are also distinguished by ViDa, suggesting that our embeddings are biophysically meaningful.

\section{Conclusion}
In this work we present ViDa, a visualization tool for DNA reaction trajectories. 
It embeds DNA secondary structures emitted by elementary-step reaction simulators in a 2D landscape, using semi-supervised VAE embedding that leverages domain knowledge to determine custom training loss terms. With two well-studied DNA hybridization reactions, we show how Vida can visually cluster trajectory ensembles into reaction mechanisms, therefore making simulation results more interpretable. ViDa also supports interactive exploration of the landscape and trajectories (details not included).

In the context of multi-stranded reactions, an important direction for improving our method is the partitioning of secondary structure microstates into clusters corresponding to different strand-level complexes, i.e., into macrostates defined by the subset of available strands which are actually bound into a complex.
For our simple example of DNA hybridization, states without inter-strand base pairs (dissociated states) should ideally be separated from those with inter-strand base pairs (associated states). For reactions involving three strands, such as three-way strand displacement, there should be 5 distinct groups (1 group without inter-strand base pairs, 3 groups with a single dissociated strand each, and 1 group with the three-way complex). However, ViDa's embeddings currently do not provide such a separation of groups for our DNA hybridization reaction samples (see Figure \ref{fig:pairunpair} in Appendix \ref{connectunconnect}). Further work will be undertaken to address this limitation.

In future work, we plan to overlay our embedding for Gao-P4T4 with structural types to gain more specific insight into its reaction mechanisms. 
We also plan to generalize ViDa to three-way strand displacement reactions, as well as to RNA reactions, such as those studied by Castro et al. \cite{GSAE}, since there are some discrepancies between their visualizations and the experimental results. Furthermore, it would be useful if trajectory samples could be classified automatically according to their time (e.g. fast) and probability (e.g. rare) to understand the contribution of individual energy basins to the overall kinetics.

\section*{Reproducibility}
The VAE was trained on an Apple M1 Pro with a 10-core CPU, 14-core GPU, 16-core Neral Engine, and 32 GB RAM. Our code is available at the GitHub repository \url{https://github.com/chenwei-zhang/ViDa}.

\section*{Acknowledgments and Disclosure of Funding}
This work has been supported by NSERC Discovery Grant. The authors thank Erik Winfree for his insightful feedback.

\newpage

\bibliographystyle{unsrt}
\bibliography{bibliography}

\clearpage

\appendix

\section*{Appendices}

\section{The Multistrand simulator} \label{multistrand}
Multistrand is a coarse-grained CTMC model designed to simulate thermodynamic and kinetic process for various DNA or RNA-strand interactions ignoring formation of pseudoknotted structures. As the name suggests, Multistrand is able to handle systems involving several distinct strands. 
Because the secondary structure state space is known to scale exponentially in the length of the strands, the simulator uses a Gillespie sampling approach, rather than representing the entire state space of secondary structures explicitly. Transitions between neighbour states are based on elementary steps, i.e., a single base pair forming or breaking. The rates between adjacent states are determined by a kinetic model, which is chosen in a way that detailed-balance is satisfied, and that the equilibrium state distribution is in line with thermodynamic predictions made by both NUPACK \cite{nupack} or Vienna RNA \cite{vienna} models.
The outputs from Multistrand include a sequence of secondary structures represented by the dp notation, the reaction simulation time (in terms of sampled trajectory time, not wall-clock time) and the corresponding free energy of the secondary structure. 

There are several simulation modes in Multistrand. The simplest one is ``trajectory mode'' which was used for the reaction from Gao et al. In this mode, we collect reactive trajectory samples (form final double-helix structure). The other one used for the reaction from Hata et al. is ``first step mode''. With this mode, we assume every Markov simulation begins with an initial ``join'' step, i.e. a pair of molecules A and B interact and form a single base pair. Therefore, the initial structure is not deterministic. In this mode, we collect both reactive and non-reactive (connected strands disassociate to separate ones during reaction proceeding) trajectory samples.

\section{Dot-parenthesis notation} \label{dp}
Dot-parenthesis (dp) notation is a simple way to represent a secondary structure of DNA or RNA. Each character represents a base (except ``\&'' and ``+'', which are separators for different strands). Dots indicate unpaired bases and matching parentheses indicate paired bases. The number of open and closed parentheses is always equal. For example, in the dp notation  3$'$-{...(((...}-5$'$+3$'$-{...)))...}-5$'$ for the secondary structure of two DNA strands $A$ (3$'$-TGACGATCA-5$'$) and $\bar{A}$ (3$'$-TGATCGTCA-5$'$), the left part of the ``+'' sign corresponds to strand $A$ and the right part corresponds to strand $\bar{A}$. Three open parentheses indicate that the bases ``CGA'' in strand $A$ are paired with the bases ``TCG'' in strand $\bar{A}$ which are represented by three closed parentheses.

\clearpage

\section{Estimated minimum passage time} \label{mfpt}

$d_{i,j}$ was obtained as the shortest path length between nodes $i,j$, using Dijkstra's algorithm \cite{dijkstra}, on a weighted undirected graph which was constructed in a pre-processing stage from the simulated Multistrand trajectories. In particular, two secondary structure nodes are connected in this graph if at least one of the two possible directions was observed in the training dataset of \emph{elementary transitions}. The edge weight was then chosen to represent the minimum expected holding time between the two adjacent states, where the expected holding time for each state was estimated as the empirical average of the sampled outgoing transition times.

\section{Stack, mis-stack, and hairpin schematic representations} \label{hairpin}

\renewcommand{\thefigure}{A\arabic{figure}}
\setcounter{figure}{0}

\begin{figure}[h]
    \centering
    \begin{subfigure}[b]{0.49\textwidth}
        \includegraphics[width=1\linewidth, trim={0cm 12cm 16cm 1cm} ,clip] 
        {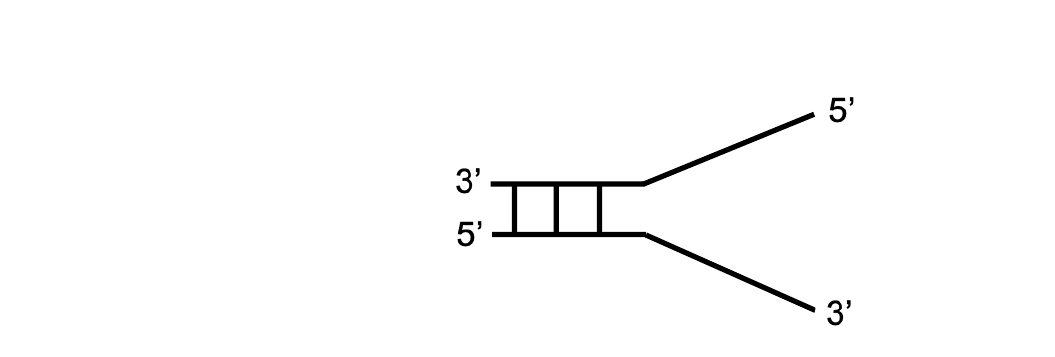}
        \caption{Stack (size 3) \label{fig:1stack}}
    \end{subfigure}
    \begin{subfigure}[b]{0.5\textwidth}
        \includegraphics[width=1\linewidth, trim={16cm 12cm 0cm 1cm} ,clip]
        {Figures/hairpin_schematic.pdf}
        \caption{Mis-stack (size 3) \label{fig:1mis-stack}}
    \end{subfigure}\hfil
    \begin{subfigure}[b]{0.49\textwidth}
        \includegraphics[width=1\linewidth, trim={0cm 5cm 16cm 7cm} ,clip] 
        {Figures/hairpin_schematic.pdf}
        \caption{4-stem hairpin \label{fig:4stem}}
    \end{subfigure}
    \begin{subfigure}[b]{0.5\textwidth}
        \includegraphics[width=1\linewidth, trim={16cm 5cm 0cm 7cm} ,clip]
        {Figures/hairpin_schematic.pdf}
        \caption{3-stem hairpin \label{fig:3stem}}
    \end{subfigure}\hfil
    \caption{ 
    Schematic representations of stack, mis-stack, and hairpin structures.
    	}
    \label{fig:hairpins}
\end{figure}

\clearpage

\section{DBSCAN} \label{DBSCAN}

\begin{figure}[h]
  \centering
  \includegraphics[width=1\linewidth,trim={8cm 4cm 8cm 3cm} ,clip]
  {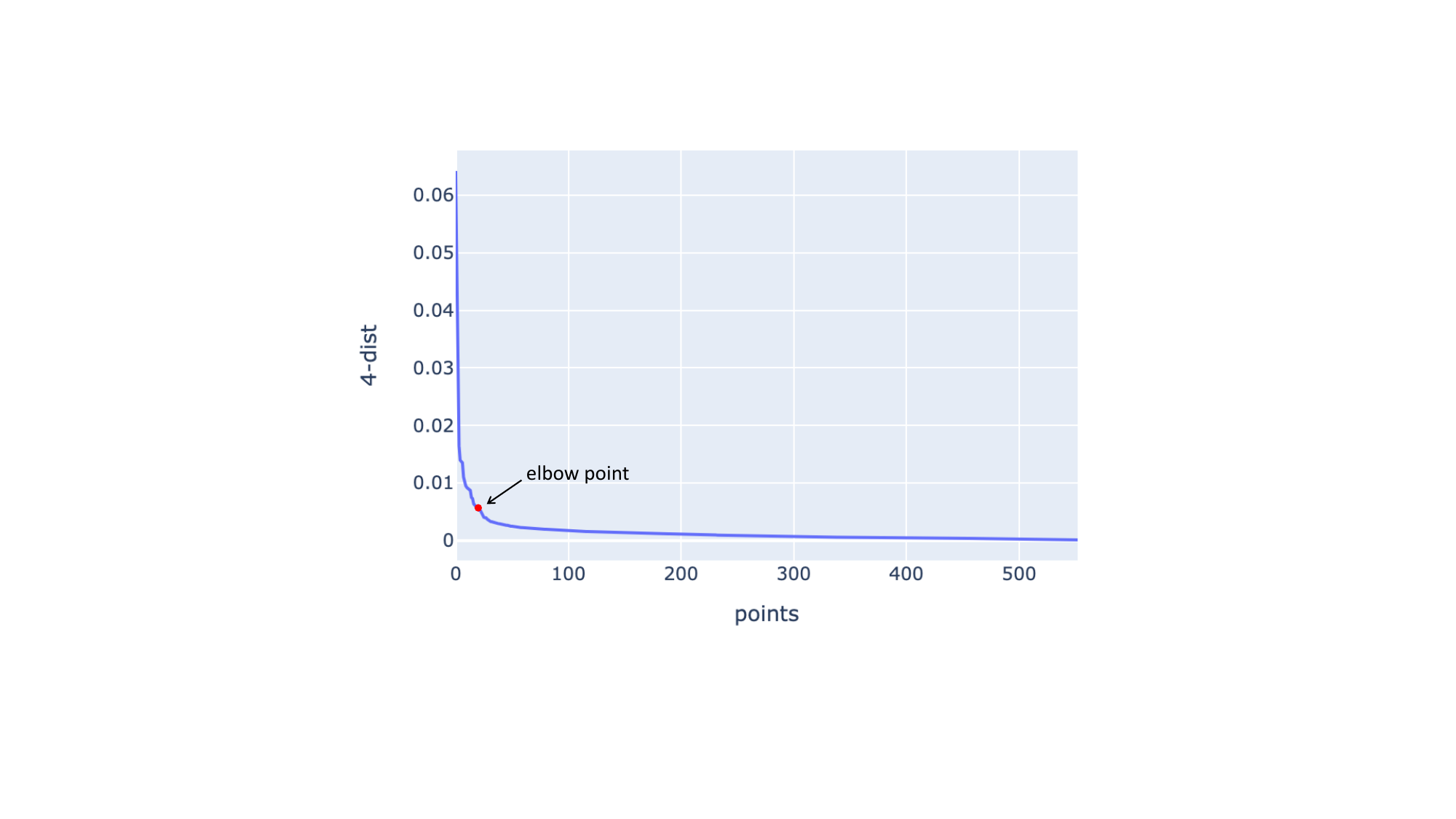}
  \caption{The elbow method plot to determine the epsilon of DBSCAN.}
  \label{fig:dbscan}
\end{figure}

DBSCAN (Density-Based Spatial Clustering of Applications with Noise) is a density-based clustering algorithm widely used in unsupervised learning. It groups data points based on their proximity in a feature space, making it particularly effective for discovering clusters of arbitrary shape. DBSCAN has two hyperparameters: epsilon ($eps$) that defines the maximum distance between two points for one to be considered as in the neighbourhood of the other, and  minimum samples ($min\_samples$) that presents the number of neighbours needed to tell a region is dense. In our work, we chose $min\_samples=4$ as the paper suggests \cite{DBSCAN}. We used the ``elbow'' method to determine $eps$. Specifically, we computed the distance of each point to its 4 nearest neighbours then sorted the points based on the resulting distances. The distances are plotted against sorted points in Figure \ref{fig:dbscan}. Finally, we selected the elbow point represented by a red dot as a reference. Therefore, the value of $eps$ is set from the reference distance, i.e. $eps=0.005$ in this work.

\clearpage

\section{Visualizations by PCA, PHATE, GSAE+PCA, GSAE+PHATE, and MDS} \label{otherviz}

\begin{figure}[h]
    \centering
    \begin{subfigure}[b]{0.49\textwidth}
        \includegraphics[width=1\linewidth, trim={0cm 0cm 17cm 9.7cm} ,clip] 
        {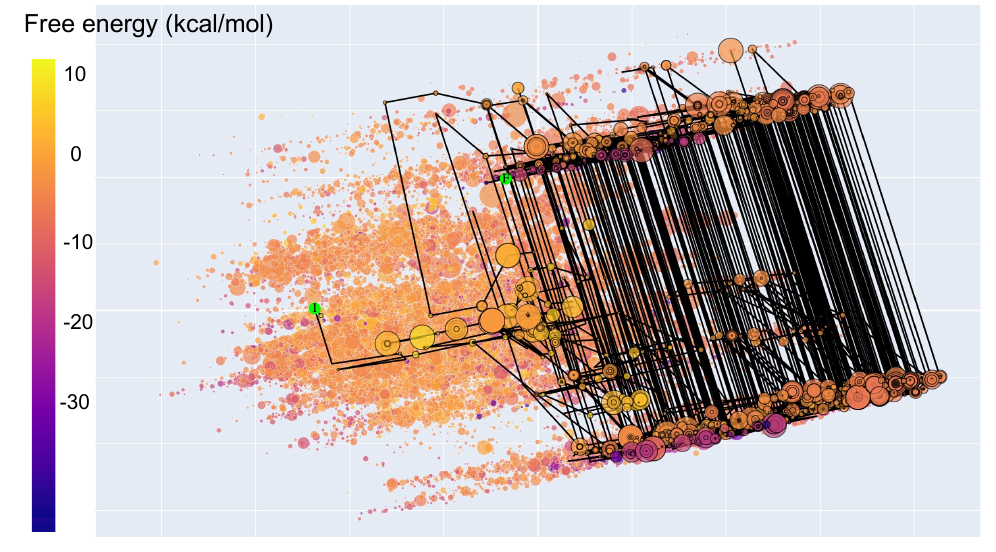}\hfil
        \caption{PCA for Gao-P4T4 \label{fig:pca}}
    \end{subfigure}
    \begin{subfigure}[b]{0.5\textwidth}
        \includegraphics[width=1\linewidth, trim={17cm 0cm 0cm 10cm} ,clip]
        {Figures/novida.pdf}
        \caption{PHATE for Gao-P4T4 \label{fig:phate}}
    \end{subfigure}\hfil
    \begin{subfigure}[b]{0.49\textwidth}
        \includegraphics[width=1\linewidth, trim={0cm 9.5cm 17cm 0cm} ,clip] 
        {Figures/novida.pdf}\hfil
        \caption{GSAE+PCA for Gao-P4T4 \label{fig:gsaepca}}
    \end{subfigure}
    \begin{subfigure}[b]{0.5\textwidth}
        \includegraphics[width=1\linewidth, trim={17cm 9.7cm 0cm 0cm} ,clip]
        {Figures/novida.pdf}
        \caption{GSAE+PHATE for Gao-P4T4 \label{fig:gsaephate}}
    \end{subfigure}
    \begin{subfigure}[b]{0.5\textwidth}
        \includegraphics[width=1\linewidth, trim={5.5cm 5cm 11cm 4cm} ,clip]
        {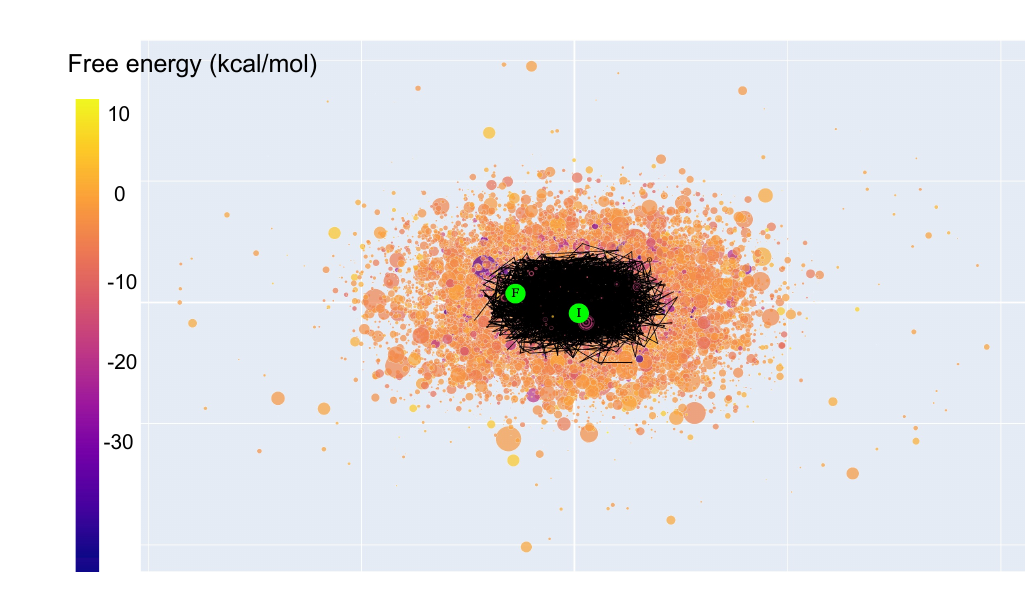}\hfil
        \caption{MDS for Gao-P4T4 \label{fig:mds}}
    \end{subfigure}
    \caption{ 
    Trajectories laid out on the embedding for Gao-P4T4. Each point represents a secondary structure state. The colour of each point represents the value of free energy. The black curve represents a trajectory. The initial and final states are indicated by the green circles marked $I$ and $F$, respectively.
    The plot made by \textbf{(a)} PCA, \textbf{(b)} PHATE, \textbf{(c)} GSAE+PCA, \textbf{(d)} GSAE+PHATE, and \textbf{(e)} MDS.
    	}
    \label{fig:novida}
\end{figure}

We assessed the visualizations for Gao-P4T4 by PCA, PHATE, GSAE+PCA, and GSAE+PHATE, as shown in Figure \ref{fig:novida}. It can be seen that PCA, PHATE, and MDS's visualizations are significantly poor. Figure \ref{fig:gsaepca} and Figure \ref{fig:gsaephate} show that energy landscapes follow high-to-low trend from the initial to final states, which is to be expected as GSAE takes the energy as a part of features. However, investigating a specified state and its neighbours, we find a large variability of their secondary structures, revealing the failure of preserving local structure. Moreover, their trajectories are not as smooth as ViDa's either owing to the large number of long segments.

\clearpage

\section{ViDa's visualization for connected and unconnected secondary structures} \label{connectunconnect}

\begin{figure}[h]
  \centering
  \includegraphics[width=1\linewidth,trim={2cm 3cm 4.5cm 2cm} ,clip]
  {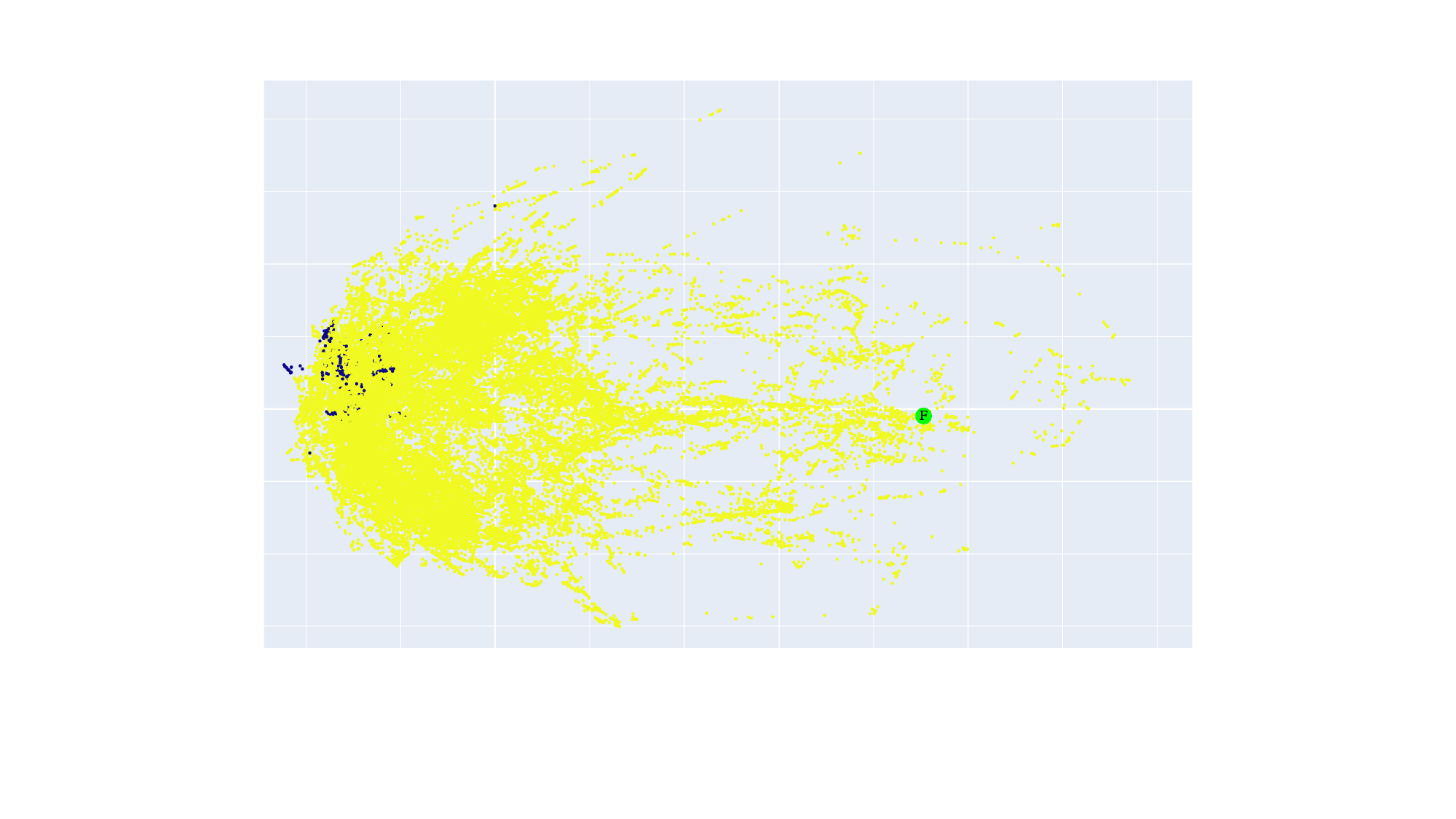}
  \caption{Visualization of the state space of connected and unconnected secondary structures for Hata-39. The points represented in dark blue and yellow refer to unconnected and connected structures, respectively. The final state is indicated by the green circle marked $F$.}
  \label{fig:pairunpair}
\end{figure}

In Figure \ref{fig:pairunpair}, connected secondary structures (i.e., those with at least one inter-strand base pair) are depicted in yellow, while structures with two single-stranded components are depicted in dark blue. It can been seen that some yellow and dark blue points are overlapped, which is not ideal. Domain experts would appreciate a dimensionality reduction method that distinguishes between states with inter-strand base pairs from those with no such pairs, keeping them separate from each other. Generalizing our methods when there are multiple interacting strands, and thus many different possible connected components involving different subsets of the strands, presents an interesting research challenge.

\end{document}